\documentclass[aps,prev,twocolumn,preprintnumbers,floatfix,nofootinbib]{revtex4-1}
\pdfoutput=1
\usepackage{graphicx}
\usepackage{bm}
\usepackage{times}
\usepackage{slashed}
\usepackage{color}
\usepackage{aas_macros}
\usepackage{slashed}
\usepackage{lipsum}
\usepackage{subfigure}
\usepackage{multirow}
\usepackage{amsmath}
\usepackage{array} 
\usepackage{varwidth} 

\newcommand{\be}{\begin{equation}}
\newcommand{\ee}{\end{equation}}
\newcommand{\bea}{\begin{eqnarray}}
\newcommand{\eea}{\end{eqnarray}}

\begin{document}

\title{Putting Things Back Where They Belong: Tracing Cosmic-Ray Injection with H$_2$}
\author{Eric Carlson}
\email{erccarls@ucsc.edu}
\affiliation{Department of Physics and Santa Cruz Institute for Particle
Physics, University of California, Santa Cruz, CA 95064, USA}

\author{Tim Linden}
\email{linden.70@osu.edu}
\affiliation{Kavli Institute for Cosmological Physics, University of Chicago, IL 60637, USA}
\affiliation{Center for Cosmology and AstroParticle Physics (CCAPP) and Department of Physics, The Ohio State University Columbus, OH, 43210 }

\author{Stefano Profumo}
\email{profumo@ucsc.edu}
\affiliation{Department of Physics and Santa Cruz Institute for Particle
Physics, University of California, Santa Cruz, CA 95064, USA}

\begin{abstract}
At present, all physical models of diffuse Galactic $\gamma$-ray emission assume that the distribution of cosmic-ray sources traces the observed populations of either OB stars, pulsars, or supernova remnants. However, since H$_2$-rich regions host significant star formation and numerous supernova remnants, the morphology of observed H$_2$ gas should also provide a physically motivated, high-resolution tracer for cosmic-ray injection. We assess the impact of utilizing H$_2$ as a tracer for cosmic-ray injection on models of diffuse Galactic $\gamma$-ray emission. We employ state-of-the-art 3D particle diffusion and gas density models, along with a physical model for the star-formation rate based on global Schmidt laws. Allowing a fraction, f$_{H_2}$, of cosmic-ray sources to trace the observed H$_2$ density, we find that a theoretically well-motivated value $f_{H_2}\sim\ $0.20 -- 0.25 (i) provides a significantly better global fit to the diffuse Galactic $\gamma$-ray sky and (ii) highly suppresses the intensity of the residual $\gamma$-ray emission from the Galactic center region. Specifically, in models utilizing our best global fit values of $f_{H_2}\sim\ $0.20 -- 0.25, the spectrum of the galactic center $\gamma$-ray excess is drastically affected, and the morphology of the excess becomes inconsistent with predictions for dark matter annihilation.
\end{abstract}

\maketitle

Observations with the Fermi Large Area Telescope (Fermi-LAT) indicating the existence of a $\gamma$-ray excess in the central regions of the Milky Way Galaxy have garnered significant interest, as they could be related to the first non-gravitational manifestation of the dark matter thought to dominate the matter content of the universe~\citep{Goodenough:2009gk, Daylan:2014rsa}. As the Fermi-LAT continues to deliver more and more accurate maps of the $\gamma$-ray sky, the existence and interpretation of this excess becomes increasingly dependent on the quality of astrophysical diffuse $\gamma$-ray foreground models. In the central regions of the Galaxy, such emission is dominated by interactions of Galactic cosmic-rays with the interstellar medium, as well as by unresolved $\gamma$-ray sources.

Physical predictions for the Galactic diffuse $\gamma$-ray emission rely on modeling Galactic cosmic rays, including their injection, propagation, and energy losses. These models hinge on sophisticated diffusion algorithms, such as {\tt Galprop} \cite{galprop0,galprop1,galprop2} or {\tt Dragon} \cite{2008JCAP...10..018E,2013PhRvL.111b1102G}, which have typically employed two-dimensional treatments of cosmic-ray propagation and of the Galactic gas density distribution. In addition, these codes utilize limited and, as we will argue below, unphysical choices for the spatial distribution of injected cosmic rays.

In the present study, we advance the state of the art for models of the Galactic diffuse emission in three key directions:

(1) A 3-dimensional treatment of cosmic-ray propagation;

(2) Up-to-date 3-dimensional gas density models;

(3) Physical models for the cosmic-ray injection morphology, a fraction of which will be posited to trace the observed H$_2$ gas density.

To accomplish the three tasks above we have suitably modified the {\tt Galprop} code \cite{galprop0,galprop1,galprop2}, and compared our predictions with 6.9 years of data from Fermi-LAT's most recent Pass 8 data release\footnote{All photons from {\tt P8R2\_CLEAN}, in a time range MET = 239557447 -- 456959280 and using standard analysis cuts.}. Additional details will be given in forthcoming publications~\cite{paper2,paper3}.

Supernova remants (SNR) are widely considered to be the dominant acceleration sites for Galactic cosmic rays \cite{Blasi2013}. Current physical models for the diffuse Galactic gamma-ray emission assume cylindrically symmetric, time-independent cosmic-ray sources, with a universal injection spectrum. The radial distribution of sources is posited to follow the observed distribution of SNR directly \cite{Case:1998}, or of SNR tracers such as pulsars \cite{Lorimer:2004, Lorimer:2006, Yusifov:2004} or OB star-forming regions \cite{Bronfman:2000}. Tracers are employed because they offer greatly improved statistics and distance estimates. However, in the Galactic center region, all currently utilized distributions are plagued by a variety of selection effects and analysis complications. For example, the assumed free electron density greatly impacts pulsar distance measurements, and unreliable star formation tracers affect the use of observed OB regions~\cite{Bally:1987}. The distribution of cosmic-ray sources is then calculated using simplified functional forms which set the density to zero at the center of the Galaxy in all but one instance \cite{Yusifov:2004}. Thus, current models completely neglect cosmic rays originating from one of the Galaxy's most extreme and SNR-dense environments, the Central Molecular Zone (CMZ). Overall, the posited cosmic-ray source distributions employed thus far in large-scale models of the Galactic diffuse emission are (i) systematically and artificially under-abundant in the central regions, and (ii) unreflective of important morphological structures such as the central bar and spiral arms, which are lost in the azimuthal average.

Here, we exploit the well-known connection between supernovae and star-forming regions \cite{Montmerle1979,Montmerle2009}, and hypothesize that a fraction $f_{\rm H2}$ of cosmic rays are injected with a spatial distribution tracing the density of collapsed $\rm H_2$ molecular clouds, with the remaining fraction $(1-f_{\rm H2})$, reflecting ``older'' cosmic rays, distributed according to the traditional axisymmetric distribution of SNR. This model is theoretically well-motivated, because high-mass OB stars, the predecessors to Type II supernovae, evolve on time scales 2-4 times shorter than the 15-20 Myr lifetime of giant molecular clouds~\cite{2011ApJ...729..133M}. This implies that a significant fraction of Galactic cosmic rays should be produced within observed star-forming regions. We employ high-resolution ($\sim$100 pc) three-dimensional $\rm H_2$ density maps that utilize gas flow simulations to resolve non-circular velocities in the inner Galaxy~\cite{PEB}\footnote{In this {\em Letter}, we use the new gas models only for generating secondary species and distributing cosmic-ray sources.  Their use for $\gamma$-ray generation does not significantly impact the conclusions here and is explored in detail in a forthcoming publication~\cite{paper3}.}, and a simple model for the star formation rate $\dot\rho_*\propto \rho_{\rm gas}^{1.5}$ \cite{1959ApJ...129..243S, 1959ApJ...129..243S}. We additionally assume a critical gas density $\rho_c=0.1\ {\rm cm}^{-3}$ under which star formation, and thus cosmic-ray acceleration, is shut off. The cosmic-ray injection intensity tracing the $\rm H_2$ gas density is calculated as:

\begin{eqnarray}
Q_{\rm CR}(\vec r) \propto
\begin{cases} 
      0 & \rho_{\rm H2} < \rho_c; \\
      \rho_{\rm H2}^{1.5} & \rho_{\rm H2} \geq \rho_c. \\ 
   \end{cases}
\label{eqn:SFR_source_model}
\end{eqnarray}

Of course, the gas density distribution measured at the present time does not reflect the distribution of cosmic-ray sources at past epochs, which is why we assume a $(1-f_{\rm H2})$  fraction of ``older'' cosmic rays to be distributed according to the axisymmetric SNR prescription. Diffusion and the rotation of the inner Galaxy largely wash out the structure of cosmic-rays on timescales shorter than the typical residence time of Galactic cosmic-ray nuclei ($\tau_{\rm res}\approx 10^7-10^8$ Myr~\cite{Lipari:2014zna}), physically motivating values of $f_{\rm H2}\gtrsim 0.1$. We also studied the effect of changing the Schmidt power-law index $n_s$ and the critical density $\rho_c$ from the default values employed here. We find that, barring extreme scenarios, the impact of these parameters is subdominant compared to $f_{\rm H2}$ \cite{paper2} and does not strongly affect the results we summarize below.

\begin{figure}[tbp]
  \centering
  \subfigure{
    \centering
  \includegraphics[width=.45\textwidth]{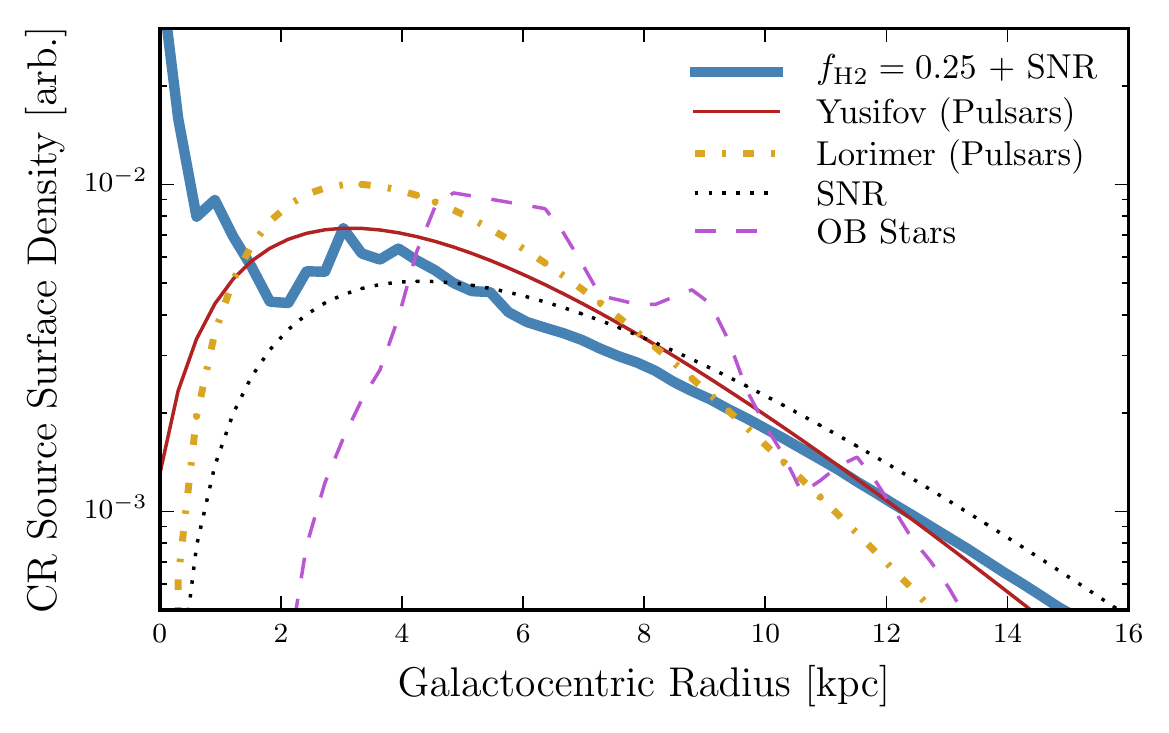}
}\\ \vspace*{-1.2em}
  \subfigure{
  \centering
   \includegraphics[width=.45\textwidth]{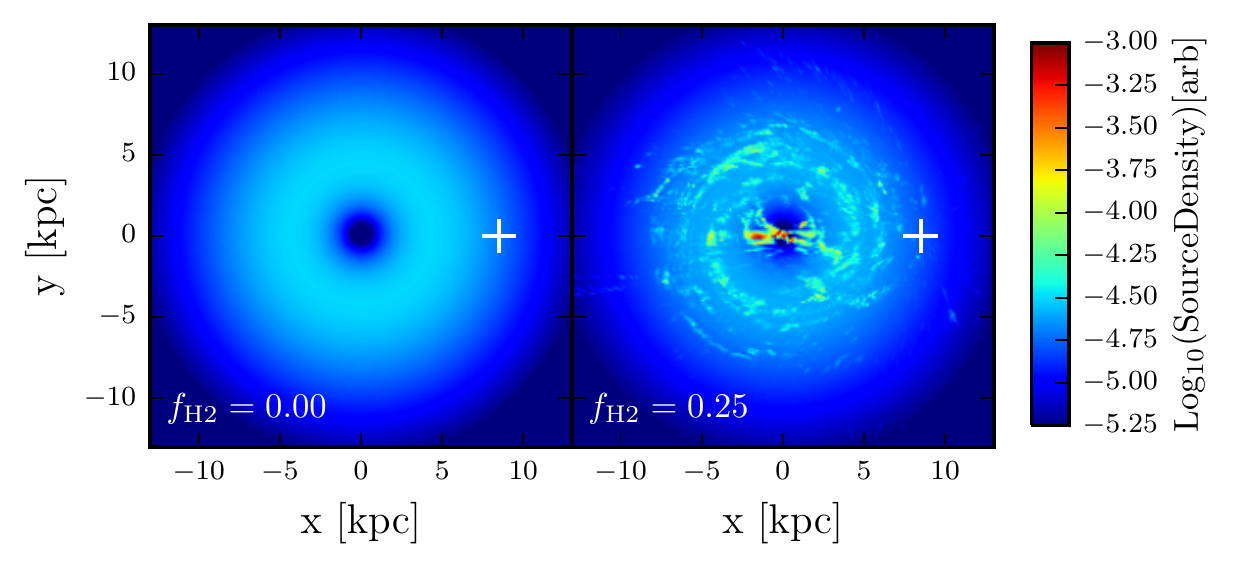}
}
\caption{  {\bf Top:} The azimuthally averaged surface density of cosmic-ray source distributions utilizing our new 3D model shown in thick blue, compared to  the traditional axisymmetric models based on SNR, pulsars, and OB stars. {\bf Bottom:} Face-on view of the cosmic-ray source surface density for the traditional SNR distribution (left) and for the best-fit star formation model, $f_{\rm H2}=.25$, (right).  The solar position is indicated with the `+' symbol.}
\label{fig:CR_sources}
\end{figure}

In the top panel of Figure~\ref{fig:CR_sources}, we compare the commonly-employed choices for the azimuthally-averaged surface density of cosmic-ray sources with a model where a fraction  $f_{H2}=0.25$ of cosmic-ray sources are embedded in H$_2$ regions according to the prescription outlined above. As we discuss below, $f_{H2}=0.2-0.25$ corresponds to the best global fit to the Fermi-LAT diffuse $\gamma$-ray sky. The bottom panels show a face-on view of the source density for the SNR model (corresponding to $f_{H2}=0$) and for the $f_{H2}=0.25$ model. Figure~\ref{fig:CR_sources} dramatically highlights the unphysical scarcity of cosmic-ray sources in the innermost kiloparsec of the Galaxy. While we note that the {\em present} rate of star formation in the CMZ is observed to be suppressed compared with that predicted via the Kennicutt-Schmidt law~\cite{Kruijssen2014}, significant multiwavelength evidence points to episodic starburst on the $\mathcal{O}(\rm Myr)$ timescales relevant here~\cite{Krumholz:2015}, with a significant event ocurring $\sim$6 Myr ago, near the lifetime of massive OB stars.  Throughout this paper, we assume a constant injection until the present day, although time-dependent effects may play a significant role~\cite{Carlson:2014,Petrovic:2014,Cholis:2015}. In addition to the CMZ, a gas-rich bar is present along the Galactic center line-of-sight (see Figure~\ref{fig:CR_sources}), which enhances cosmic-ray sources toward the Galactic center, a feature otherwise lost using a cylindrically-symmetric treatment.

As will be discussed in detail in forthcoming publications \cite{paper2,paper3}, the addition of a cosmic-ray injection source distribution tracing H$_2$ gas has a net effect on the steady-state GC cosmic-ray density (after propagation) of nearly one order of magnitude. This enhancement is especially dramatic for cosmic-ray electrons, where the density remains larger than a factor of two out to nearly 5~kpc from the GC. Notably, the local cosmic-ray density is essentially unaffected.

\begin{figure}[tbp]
\centering
\includegraphics[width=.4\textwidth]{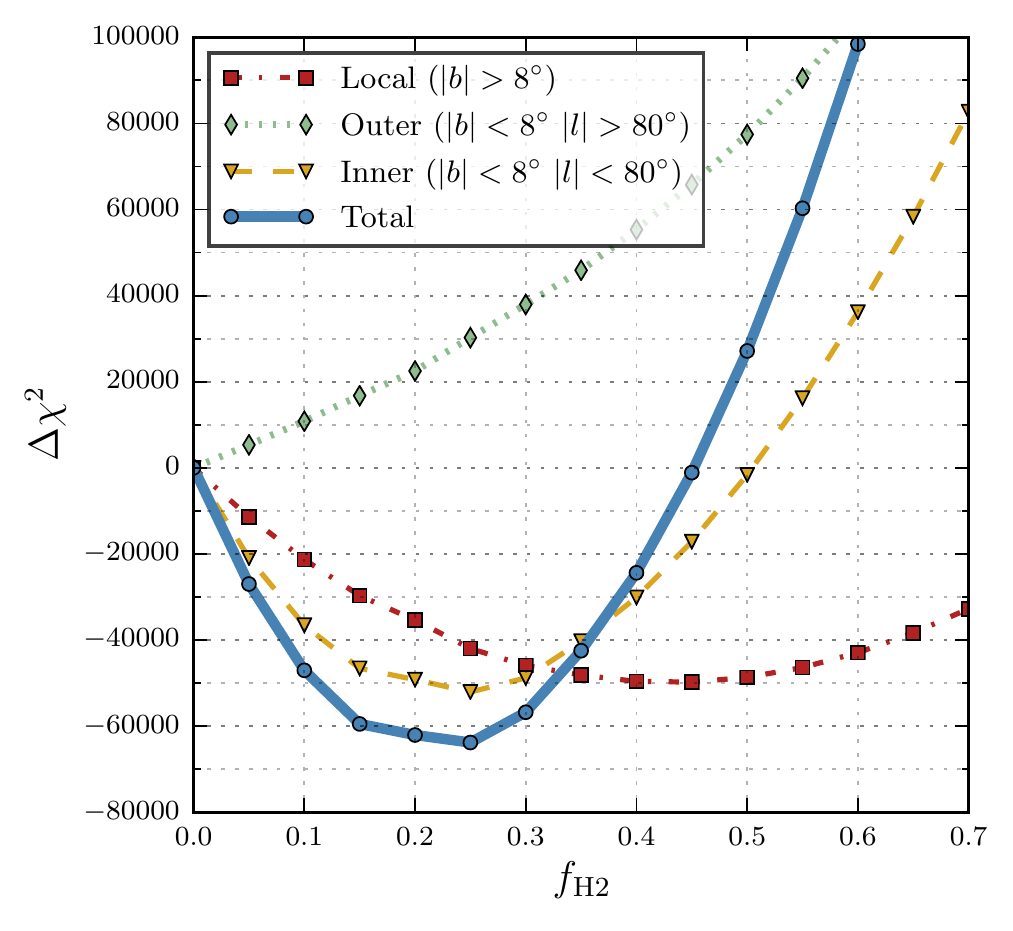}
\caption{ $\Delta \chi^2$ as a function of $f_{\rm H2}$ for several regions of the global $\gamma$-ray analysis.}
\label{fig:delta_chisq}
\end{figure}

While our model is physically well motivated, it is paramount to assess whether a non-zero value for $f_{H2}$ yields a better or worse fit to the diffuse $\gamma$-ray sky overall. We perform a `Global' binned likelihood analysis in three regions of the Galaxy: inner ($|l|<80^\circ, |b|<8^\circ$), outer ($|l|>80^\circ, |b|<8^\circ$), and local ($|b|>8^\circ$). Our adopted statistical framework, point source masking, photon binning ($\approx .23^\circ$ pixels in 24 energy bins), and extra templates (isotropic~\cite{isotropic} + Fermi Bubbles~\cite{Su:2010qj}) are identical to those used in Ref.~\cite{Calore:2015}. As $f_{\rm H2}$ is increased, cosmic rays are redistributed through the Galaxy, and we allow for radial variations in the $\rm CO \to H_2$ conversion factor using 9 Galactocentric rings~\cite{fermi_diffuse}. In these preliminary fits the spectrum of the diffuse components in the {\em Global} analysis is fixed in order to limit the number of degrees of freedom. Each point source is adaptively masked and fixed to its 3FGL flux and spectrum~\cite{3FGL}.

In Figure~\ref{fig:delta_chisq} we plot the log-likelihood of our model fit to the diffuse $\gamma$-ray emission as a function of $f_{H2}$, compared to a baseline model of $f_{H2}=0$, i.e. with cosmic-ray sources distributed according to the  axisymmetric SNR model. In the inner and local regions, turning on cosmic-ray sources in H$_2$ regions dramatically improves the quality of the global fit to the observed diffuse emission\footnote{Although the value of $\Delta \chi^2$ in the outer galaxy becomes monotonically worse, this region is metal-poor such that the $\rm H_2$ density is not well traced by CO, as evidenced by unphysical preferred values of $\rm X_{CO}$ when fitting against $\gamma$-ray data in the outer Galaxy~\cite{MS:2004}.  Additionally, the total number of CR sources is constrained here, with increasing $f_{\rm H2}$ resulting in fewer sources outside the solar circle. Technical details are discussed in a forthcoming publication~\cite{paper3}.}.  The `Total' curve sums all three regions, showing an optimal fraction $f_{H2}\simeq 0.25$ overall, with the local region preferring even higher values up to $f_{H2}\simeq 0.45$.  Examining the pixel-by-pixel $\Delta\chi^2$ of each region reveals that the `local' improvements are most significant near the disk and especially for $-10^\circ<l<30^\circ$ where cosmic-rays from the bar and inner molecular arms illuminate the interstellar medium.  For the `inner' region, $|l|<30^\circ$ shows the most significant improvement, indicating that the new gas models are resolving important cosmic-ray emitting structures toward the inner Galaxy.  In relative terms, the new source distribution represents a genuine quantitative improvement, with a $\Delta\chi^2$ comparable to that of changing the diffusion parameters, gas distributions, or source distributions over the model space of Refs.~\cite{fermi_diffuse,Calore:2015}.

\begin{figure}[tbp]
\centering
\includegraphics[width=.48\textwidth]{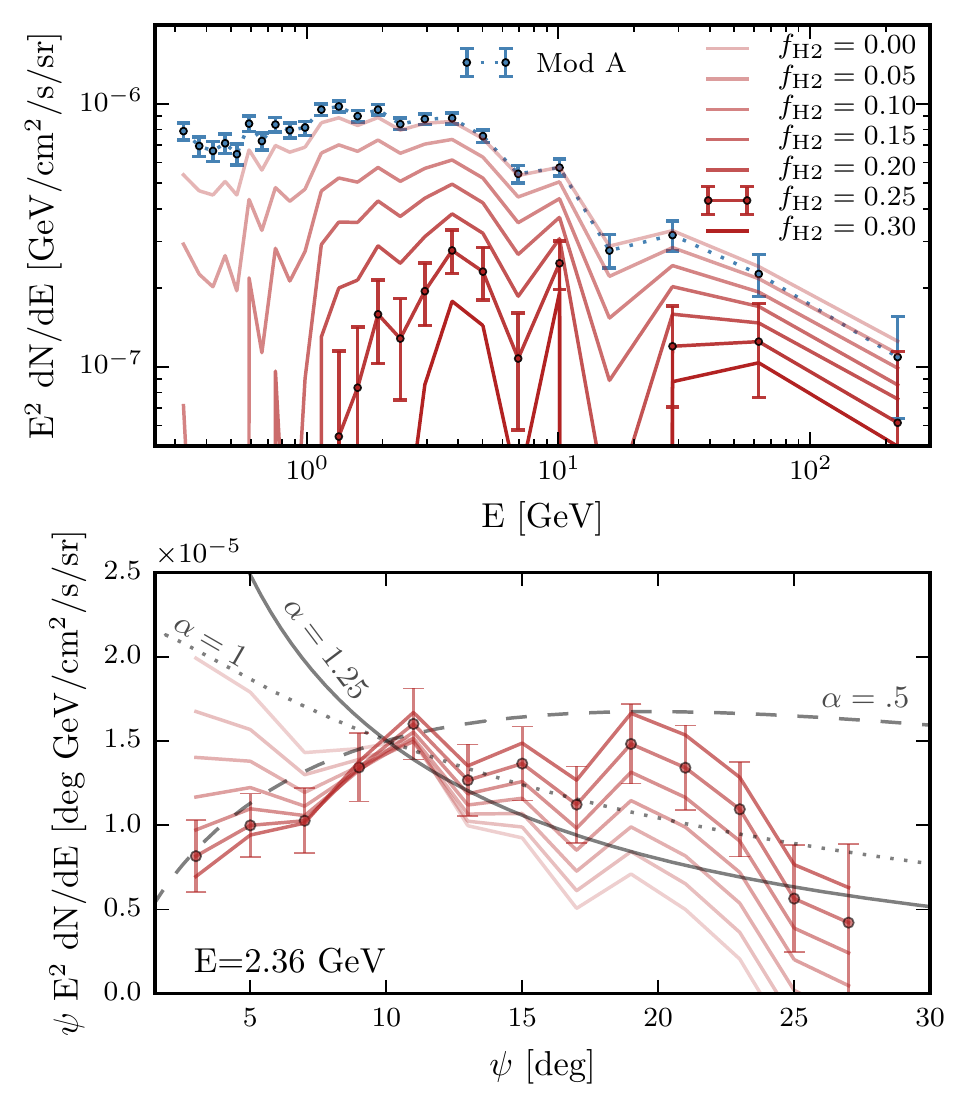}
\caption{{\bf Top} Spectrum of the Galactic center `excess' as $f_{\rm H2}$ is increased in increments of 0.05 (light-to-dark red).  We also show the spectrum and statistical error-bars of the benchmark Mod A from Ref.~\cite{Calore:2015} (blue). {\bf Bottom:} Flux of the Galactic center excess as a function of the angle from the Galactic center for the peak energy bin. Also shown are projected power-law profiles for the three-dimensional $\gamma$-ray emission intensity, which are equivalent to the square of the corresponding three-dimensional dark matter density distribution. }
\label{fig:spectrum_radial}
\end{figure}

The addition of cosmic-ray sources in star-forming regions strongly affects the prediction for the diffuse astrophysical $\gamma$-ray emission in the Galactic center region. It is thus paramount to ascertain how this affects the properties of the claimed Galactic center excess~\citep{Daylan:2014rsa}. We use the analysis framework described above on a new region of interest, the 
{\em Inner Galaxy}, defined by $|l|<20^\circ, 2^\circ<|b|<20^\circ$, noting that the bright Galactic plane is masked in order to probe the extended properties of the excess.  To evaluate the spectrum and intensity of the $\gamma$-ray excess, we add an additional template with a morphology calculated using a generalized NFW profile~\citep{Navarro:1996gj} with an inner slope $\alpha$~=~1.25. For each value of f$_{H_2}$ we allow the normalization of the NFW profile, diffuse models, isotropic models, and Fermi Bubbles to float independently in each energy bin, fixing only point sources to their 3FGL values.

In the upper panel of Figure~\ref{fig:spectrum_radial} we show the spectral properties of the NFW template in the Galactic center vicinity for increasing values of $f_{H2}$, and compare with the baseline Mod A of Ref.~\cite{Calore:2015}. The effect on the central gamma-ray excess is dramatic: an increasing fraction of cosmic rays injected in H$_2$ regions yields a substantial suppression of the excess across all energies. The effect is most dramatic at lower energies, where the suppression of the excess emission is larger than an order of magnitude,
but it continues into the GeV energy range and is consistently larger than a factor of 2 for the values of $f_{H2}\sim0.25$ preferred by the global fit analysis\footnote{Recently, Ref.~\cite{Gaggero2015} examined injecting cosmic-rays over a $200-400$ pc Gaussian CMZ, showing also a strong reduction of the Galactic center excess.  Here we also provide a concrete physical model with support from the global $\gamma$-ray sky.}. Notably, we find that the statistical significance of the NFW template is {\em maximally} reduced at the value $f_{\rm H2}\approx0.25$, which is consistent with global best-fit value.

The lower panel of Figure~\ref{fig:spectrum_radial} examines the implications for the Galactic center excess morphology.  The NFW$_{\alpha=1.25}$ template is divided into $2^\circ$ wide annuli, refitting the spectrum of each annulus simultaneously.   Shown is the flux of the annihilating dark matter template as a function of the angle from the center of the Galaxy in the peak energy bin $E=2.36$~GeV. Again, we observe a dramatic suppression of the excess emission in the central few degrees, and a general flattening of the residual emission to $\alpha=0.5$, which significantly deviates from expectations in the case of dark matter pair annihilation ($\alpha\simeq$1 -- 1.25). 

We also studied the effect of a non-zero $f_{H2}$ on the axis ratio, and found that the preferred values increases from around 1 at $f_{H2}=0$ to around 1.75 for the preferred values $f_{H2}\sim0.25$, with an axis ratio greater than one indicating elongation perpendicular to the Galactic disk.  Remarkably, if one simultaneously flattens and elongates the NFW template to the preferred values above, the normalization of the excess returns similar values as found for f$_{H2}$~=~0.0, but this flux is removed from the template tracing the Fermi bubbles. In light of this, one might consider connections between the Galactic center excess and the intersection of the low-latitude Fermi Bubbles~\cite{Su:2010qj} with the Galactic plane and/or collimated central outflows driven by intense stellar winds~\cite{Crocker:2011}.

As $f_{\rm H2}$ is increased, the central cosmic-ray population grows, most prominently for cosmic-ray electrons. This leads to a bright spherical inverse-Compton enhancement at the Galactic center which shrinks in radius with increasing energy and reduces much of the excess. Remaining `residual photons' are absorbed by both a softening isotropic spectrum below 10 GeV and a $~\sim 20\%$ enhancement to the Fermi bubbles flux above 1 GeV\footnote{Note that the spectrum of each of these templates is constrained by larger fields of view as described in Ref.~\cite{Calore:2015}.}.  

The absorption of the gamma-ray excess by physically motivated cosmic-ray injection models is stunning, especially in light of the global preference for these models in regions far from the Galactic center. Moreover, if we do not include an NFW template in our analysis of the inner galaxy region, we find a best fit value of $f_{\rm H2}\approx0.20$, which is consistent with our global analysis. However, an important caveat concerns analyses that are restricted to the inner Galaxy and also allow an NFW template to float freely in the fit. In this case, we find a statistical preference for models with $f_{\rm H2}\approx0.10$, and a gamma-ray excess with an intensity reduced by a factor of approximately 1.5 compared to its strength for $ f_{\rm H2}$~=~0. The statistical preference for this value is $\Delta \chi^2$~$\approx$~300 compared to either $f_{\rm H2}~=~0.00$ or $f_{\rm H2}~=~0.20$. Such statistical preferences are negligible compared to the preference of the global fit for $f_{\rm H2}\approx0.20$. Thus, a complete interpretation of these results for the gamma-ray excess depends on whether a high value for $f_{\rm H2}$ in the inner Galaxy is accepted as a prior based on the improved global fit, or is considered equally against emission tracing an NFW template in the Galactic center. We leave a complete analysis of this important question to forthcoming work~\citep{paper2}.

In conclusion, in this {\em Letter} we have pushed the envelope of current models for the distribution of Galactic cosmic rays in three important directions, adding: (1) a fully three-dimensional treatment of propagation, (2) a three-dimensional model of galactic gas, and (3) a variable fraction of cosmic-ray source injection tracing star-forming regions. We have discovered three important results: (i) the overall quality of the predicted diffuse Galactic $\gamma$-ray emission is significantly improved when 20-25\% of cosmic rays are injected in star-forming regions; (ii) while a Galactic center $\gamma$-ray excess persists in the inner Galaxy, the brightness of the excess is substantially reduced across all energies with the new cosmic-ray sources turned on, and (iii) the spectrum and morphology of the Galactic center excess strongly depend on the fraction of cosmic-ray sources allocated to $\rm H_2$-rich regions. Forthcoming papers will provide complete details and additional results \cite{paper2,paper3}.

\section*{Acknowledgements}
We thank Christoph Weniger, Mark Krumholz, and Andy Strong for discussions as well as Gudlaugur J\'{o}hannesson, Hiroyuki Nakanishi, and Martin Pohl for discussions and access to various datasets used in this analysis. Simulations were carried out on the UCSC supercomputer Hyades, supported by the National
Science Foundation (award number AST-1229745) and by UCSC. EC is supported by a NASA Graduate Research Fellowship under NASA NESSF Grant No. NNX13AO63H. TL is supported by the National Aeronautics and Space Administration through Einstein Postdoctoral Fellowship Award Number PF3-140110. SP is partly supported by the US Department of Energy, Contract DE-SC0010107-001.

\bibliography{biblio}

\end{document}